\documentclass[3p,times,procedia]{elsarticle}
\usepackage{nupha_ecrc}
\usepackage[nointegrals]{wasysym}
\usepackage{color}

\volume{00}

\firstpage{1}

\journalname{Nuclear Physics A}

\runauth{F. Becattini}

\jid{nupha}

\jnltitlelogo{Nuclear Physics A}

\usepackage{amssymb,amsmath}
\usepackage[figuresright]{rotating}

\newcommand{\di}{{\rm d}}

\def\spt{{\cal S}}
\def\wT{{\widehat T}}
\def\wj{{\widehat j}}

\def\wrho{{\widehat{\rho}}}
\def\wrhol{{\widehat{\rho}_{\rm LE}}}

\def\wP{{\widehat{P}}}
\def\wJ{{\widehat{J}}}
\def\wQ{{\widehat{Q}}}

\newcommand{\tr}{{\rm tr}}

\newcommand{\omegav}{\boldsymbol{\omega}}

\newcommand{\x}{{\rm x}}

\newcommand{\be}{\begin{equation}}
\newcommand{\ee}{\end{equation}}                                                                               
\newcommand{\bea}{\begin{eqnarray}}
\newcommand{\eea}{\end{eqnarray}}

\begin{document}

\begin{frontmatter}



%% Instructions from Editor: Please use the following \dochead only in the preprint version (e-print arXiv etc.); 
%% use empty \dochead{} when submitting to Nuclear Physics A!
\dochead{XXVIIth International Conference on Ultrarelativistic Nucleus-Nucleus Collisions\\ (Quark Matter 2018)}
%% e.g. \dochead{17th International Conference on Dynamical Processes in Excited States of Solids}

\title{Polarization and Chirality: the quantum features of the Quark Gluon Plasma }


\author{Francesco Becattini}

\address{Dipartimento di Fisica e Astronomia, Universit\'a di Firenze e INFN Firenze\\
Via G. Sansone 1, I-50019, Sesto Fiorentino (Firenze), Italy}

\begin{abstract}
Polarization and chirality are direct manifestations of quantum mechanics in the 
Quark Gluon Plasma as a relativistic fluid. This is one of the reasons why they
are intriguing phenomena, that have attracted so much attention lately.
In this talk I will review what is, in my view, the current theoretical 
understanding and highlight the most interesting recent result.  
\end{abstract}


\end{frontmatter}

% \linenumbers

%*******************************************************************************
\section{Introduction}
\label{intro}
%*******************************************************************************

Much effort has been recently devoted, in relativistic heavy ion physics, to the
search of chirality and polarization-related effects. The Chiral Magnetic Effect (CME) 
is yet to be discovered in heavy ion collisions but the theoretical work is now 
entering a mature stage where the endeavour for a full numerical simulation 
including dynamical electromagnetic field is ongoing. On the other hand, polarization is, 
at least on the experimental side - certainly in a more advanced stage. The first positive
evidence of a global polarization reported by the experiment STAR \cite{starnature}
has been confirmed by more and more accurate measurements that have been presented 
in this conference \cite{starnewpol,niida}. This result has generated much enthusiasm 
not just among those working on the subject, but also among those interested in chiral-
related effects. Indeed, these two phenomena have several common features: they are 
both related to the polarization degrees of freedom, they should arise only in 
peripheral collisions, and others. On top of that, they are direct manifestations of the 
quantum nature of the Quark Gluon Plasma (QGP); CME requires an anomaly at work, 
which is by definition a quantum breaking of a classical symmetry and polarization 
is itself a quantum object.

%*******************************************************************************
\section{The Chiral Magnetic Effect}
\label{cme}
%*******************************************************************************

I will not bring up anything about the experimental search of the CME, particularly 
about the correlators designed to obtain an evidence thereof. For that topic and 
an experimental overview on chirality and polarization I refer the reader to the talk 
by Tu in this conference \cite{kong}. 

The search for the CME is now entering - as has been mentioned in the Introduction - 
a more advanced stage where one aims at dynamically evolving both the axial charge 
and the electromagnetic field to make more accurate predictions. Supplementing the 
usual hydrodynamic model of QGP with these new quantities implies the transition from
the currently available viscous hydrodynamics code to the so-called 
Chiral-Relativistic-Magneto-Hydrodynamics ($\chi$-RMHD). 
To date, most studies have focused on the theoretical features 
of $\chi$-RMHD \cite{boyarsky,ldz,hattori} in terms of constitutive equations and all that. 
In fact, no numerical implementation is available, although there are ongoing projects
to develop it. 

While the full $\chi$-RMHD is yet to come, there have been worthwhile intermediate
steps, like ideal RMHD codes \cite{inghirami,roy} as well as a hybrid code like AFVD
(Anomalous Viscous Fluid Dynamics) \cite{avfd} wherein vector and axial currents 
are dynamicall evolved in a background electromagnetic field, while the fluid viscous 
hydrodynamics is completely decoupled. With these calculations, a charge imbalance 
for the final particles is obtained, which is of course strongly dependent on the 
initial value of the axial charge \cite{liao}.

It is worth reviewing the physics of the CME from the perspective of equilibrium 
quantum statistical mechanics.

Thermodynamic equilibrium in quantum statistical mechanics, in the grand-canonical 
ensemble, is defined through the density operator maximizing entropy $S = -\tr (\wrho \log \wrho)$
with the constraints of given mean values of {\em conserved} quantities, such as 
energy, charge etc., which leads to a density operator of the form:
\be 
 \wrho  = \frac{1}{Z} \exp[ - \widehat H/T + \mu \wQ /T ]
\ee
where $\widehat H$ is the Hamiltonian operator and $\wQ$ the conserved charge. The
parameters $1/T$ and $\mu/T$ are - as it is known - the Lagrange multipliers associated
to the constraints. For a system of free massless fermions, there is an additional 
conserved charge besides the electric/vector charge, namely the axial charge:
\be
  \wQ_A = \int \di \Sigma_\mu \; \wj^\mu_A
\ee
which is written here in a fully covariant form. The integration can be done over any
space-like 3D hypersurface because the divergence of the integrand vanishes. Therefore,
since the axial charge is conserved, one can write a density operator including 
an {\em axial chemical potential} $\mu_A$:
\be\label{densop}
 \wrho  = \frac{1}{Z} \exp[ - \widehat H/T + \mu \wQ /T + \mu_A \wQ_A /T ]
\ee
A value $\mu_A \ne 0$ signals an imbalance between right-handed and left-handed fermions.

For the mean values of general physical quantities to be non-vanishing at equilibrium, the
symmetries of the density operator (\ref{densop}) are crucial. In order for an electric 
current to be allowed at global thermodynamic equilibrium, rotational symmetry needs
to be broken along with time reversal, charge conjugation and parity. For the former 
three, a constant and uniform magnetic field is enough as it breaks them (see table~\ref{table})
in the Hamiltonian. Yet, a constant and uniform magnetic field does not break
parity; for this purpose, a non-vanishing axial chemical potential - that is a chiral 
imbalance - is necessary. This is why in the renowned formula \cite{cmes}:
$$
 {\bf j}_{el}  = \frac{e^2}{2\pi^2} \mu_A {\bf B}
$$
the coefficient $\mu_A$ plays a crucial role. 
%--------------------------------------------------------------------------------
\begin{table}[h]
\centering
\label{table}
\vspace{0.5cm}
\begin{tabular}{|c|c|c|c|}
\hline
           & B & $\mu_B$  & $\mu_A$  \\
\hline    
  C        & {\color{red}\CIRCLE}   &  {\color{red}\CIRCLE}  & {\color{green}\CIRCLE}  \\          
  P        & {\color{green}\CIRCLE} &  {\color{green}\CIRCLE}& {\color{red}\CIRCLE}    \\
  T        & {\color{red}\CIRCLE}   &  {\color{green}\CIRCLE}& {\color{green}\CIRCLE}  \\ 
  Rotation & {\color{red}\CIRCLE}   &  {\color{green}\CIRCLE}& {\color{green}\CIRCLE}   \\
\hline                
\end{tabular}
\caption{Symmetries of the density operator (\ref{densop}) which are maintained 
(green) or broken (red) by the parameters $B,\mu_B,\mu_A$.} 
\end{table}
%--------------------------------------------------------------------------------
This observation is due to Vilenkin \cite{vilenkin1}.

Indeed, the axial current is not conserved because of the anomaly:
\be\label{anoma}
\partial_\mu \wj^\mu_A = \frac{e^2}{2 \pi^2} E_\mu B^\mu = \frac{e^2}{4\pi^2} 
\partial_\mu \left( \epsilon^{\mu\lambda\rho\sigma} A_\lambda \partial_\rho A_\sigma \right) 
\ee
so one may wonder whether a global thermodynamic equilibrium analysis holds. Indeed,
this may happen under two circumstances:
\begin{itemize}
\item if the electromagnetic field is external and $E \cdot B =0$
\item if the electromagnetic field is dynamical, provided that the axial current
and the axial charge include the contribution of the Chern-Simons current reported
on the right-hand-side of eq.~(\ref{anoma}).
\end{itemize}
The second option did not receive much attention in literature, except in 
ref.~\cite{boyarsky} but one should 
never forget that the dynamical electromagnetic field is always present. In fact,
it can be shown that, in the limit of free electromagnetic field, the right hand 
side of the (\ref{anoma})
corresponds to the difference between the number of right-handed and left-handed
photons, so that one can loosely say that the total helicity of the fermion-boson
system is overall conserved \cite{buzzegoli}.
The possibility to study the Chiral Magnetic Effect in an equilibrium quantum statistical
framework is very convenient in many respects. This feature, which can be seen as 
an expression tantamount to "the CME current is non-dissipative", makes the calculation of many
quantities easier than, for instance, dissipative transport coefficients. 

It is worth stressing that, even if in this formalism the anomaly is somewhat "hidden",
it plays a crucial role to generate the chirality imbalance which is needed to 
give rise, after relaxation, to a finite $\mu_A$. Because of the non-conservation
of the axial current, a chiral charge can be created in a situation where, for some
time, an external electromagnetic field is such that $E \cdot B \ne 0$, like in 
condensed matter experiments \cite{kharnat} or in the so-called magnetic reconnection 
\cite{hirono}. In the QGP, the chiral charge is believed to arise from a topological 
non-­Abelian sphaleron transition, and quantitative theoretical predictions are 
currently being provided \cite{mace}. 

Finally, I would like to make a point about what can be called anomalous and what
should not be called anomalous. Indeed, it has been somewhat customary in literature 
to qualify as "anomalous" many non-dissipative currents which are not expected to 
appear in the traditional global thermodynamic equilibrium. For instance, the mean 
value of the axial current of the free Dirac field 
\be
 j^\mu_A = \left( \frac{T^2}{6} + \frac{\mu^2}{2\pi^2} \right) \omega^\mu
\ee
in presence of vorticity, is repeatedly defined as "anomalous". In fact, this relation 
is not of anomalous origin because it can arise in {\em any} field theory involving 
Dirac fields, simply because it is allowed by the symmetry of the density operator 
with rotation:
$$
 \wrho = \frac{1}{Z} \exp [ - \widehat H/T + \omega \widehat J_z/T ]
$$
which was noted, again, for the first time by Vilenkin \cite{vilenkin2} and recently
discussed in \cite{buzzbeca}. In my 
view, the qualification of anomalous is appropriate only when a current vanishes 
if the axial chemical potential $\mu_A$ does. The reason is explained above: without 
an anomalous divergence, it would be impossible to drive a chirality imbalance 
with external fields, that is to create a $\mu_A \ne 0$.

%*******************************************************************************
\section{Polarization}
\label{pol}
%*******************************************************************************

Particles produced in relativistic heavy ion collisions are expected to be polarized 
in peripheral collisions because of angular momentum conservation. At finite impact 
parameter, the QGP has a finite angular momentum perpendicular to the reaction plane 
and some fraction thereof may be converted into spin of final state hadrons. Therefore, 
measured particles may show a finite mean {\em global} polarization along the angular
momentum direction. In a fluid at local thermodynamic equilibrium, the polarization
can be calculated by using the principle of quantum statistical mechanics, that is
assuming that the spin degrees of freedom are at local thermodynamical equilibrium 
at the hadronization stage, much the same way as the momentum degrees of freedom.

The crucial role in the calculation of the polarization for the fluid produced in 
relativistic heavy ion collisions is played by the density operator. For a system 
at Local Thermodynamic Equilibrium (LTE), this reads \cite{betaframe}:
\be\label{gencov}
  \wrhol = (1/Z) \exp \left[- \int_\Sigma \di\Sigma_\mu  \left( \wT^{\mu\nu} \beta_\nu 
- \zeta \wj^\mu \right) \right]
\ee
where $\beta = (1/T)u$ is the four-temperature vector, $\wT$ the stress-energy tensor, $\wj$
a conserved current - like the baryon number - and $\zeta=\mu/T$. The mean value 
of a local operator $\widehat O(x)$ (such as, for instance the stress-energy tensor 
$\wT$, or the current $\wj$) at LTE:
\be
  O(x) = \tr ( \wrhol \widehat O(x) )
\ee
and if the fields $\beta$,$\zeta$ vary significantly over a distance which is much 
larger than the typical microscopic length (indeed the {\em hydrodynamic limit}), 
then they can be Taylor expanded in the density operator starting from the point 
$x$ where the mean value $O(x)$ is to be calculated. The leading terms in the 
exponent of (\ref{gencov}) then become \cite{betaframe}:
\be\label{ltedensop}
 \wrhol \simeq \frac{1}{Z_{\rm LE}}\exp \left[- \beta_\nu(x) \widehat P^\nu + 
 \xi(x) \widehat Q  - \frac{1}{4} (\partial_\nu \beta_\lambda(x) - \partial_\lambda 
 \beta_\nu(x)) \widehat J_x^{\lambda\nu} + \frac{1}{2}(\partial_\nu \beta_\lambda(x) 
 + \partial_\lambda \beta_\nu(x)) \,\widehat L^{\lambda\nu}_x + \nabla_\lambda \xi(x) 
 \, \widehat d^\lambda_x \right].
\ee
where the last two terms with the shear tensor and the gradient of $\zeta$ are 
dissipative and vanish at equilibrium. The $\nabla_\lambda$ operator stands for:
$$
  \nabla_\lambda = \partial_\lambda - u_\lambda u \cdot \partial
$$
as usual in relativistic hydrodynamics. The term which is responsible for a non-vanishing
polarization is the one involving the angular momentum-boosts operators $\widehat J_x$.

The polarization of particles in a fluid at LTE can in principle be obtained by 
calculating matrices like:
$$
 W_{\sigma,\sigma^\prime} = \tr (\wrhol a^\dagger(p)_\sigma a(p^\prime)_{\sigma^\prime})
$$
where $a(p)_\sigma$ are the destruction operators of final state particles of four-momentum
$p$ and $\sigma$ is the spin state index. Nevertheless, the exact calculation of 
$W$ is a difficult one even with the expansion of $\wrhol$ and the mean polarization
was obtained in ref.~\cite{becaspin} by means of a different method, involving
the spin tensor and an  {\em ansatz} about the form of the covariant Wigner function at
LTE (see also \cite{xnwang2}). As a result, the mean spin vector of $1/2$ particles 
with four-momentum $p$, turns out to be:
\be\label{basic}
 S^\mu(p)=  - \frac{1}{8m} \epsilon^{\mu\rho\sigma\tau} p_\tau 
 \frac{\int d\Sigma_\lambda p^\lambda n_F (1-n_F)\varpi_{\rho\sigma}}
{\int d\Sigma_\lambda p^\lambda f(x,p)}
\ee
where $n_F = (1+\exp[\beta(x) \cdot p - \mu(x) Q/T(x)] +1)^{-1}$ is the Fermi-Dirac 
distribution and $\varpi(x)$ is the {\em thermal vorticity}, that is:
\be\label{thvort}
   \varpi_{\mu\nu} = -\frac{1}{2} \left( \partial_\mu \beta_\nu - \partial_\nu 
   \beta_\mu \right)
\ee

The eq.~(\ref{basic}) has been used in all numerical calculations of polarization,
either based on the hydrodynamic model \cite{hydrocalc} or other approaches \cite{others}
and a good agreement with the data is observed. A crucial feature of the 
(\ref{basic}), and more in general of this effect, is that it predicts an almost
equal polarization of particles and anti-particles (if quantum statistics effect
are not important) for it is a statistical thermodynamic effect driven by local
equilibration and not by an external C-odd field like the electromagnetic field.
This distinctive feature is confirmed - modulo small deviations - by all measurements 
\cite{starnature,starnewpol,niida}.

While an exact derivation of the formula (\ref{basic}) is still missing, this
formula is the correct first-order expression in thermal vorticity in the non-relativistic
limit and in the limit where quantum statistics can be neglected \cite{becaspin}. 
At a glance, it appears to be the most reasonable expression of the mean spin 
vector: it is linear in thermal vorticity, the spin is orthogonal to the four-momentum
and it also has the right limit ($S^\mu=0$) for the fully degenerate Fermi gas,
i.e. $n_F=1$.

To gain insight into the physics of polarization in a relativistic fluid, it is 
very useful to decompose the gradients of the four-temperature vector in the 
eq.~(\ref{basic}). We start off with the seperation of the gradients of the comoving 
temperature and four-velocity field:
$$
 \partial_\mu \beta_\nu = \partial_\mu \left(\frac{1}{T}\right) + \frac{1}{T} 
 \partial_\mu u_\nu 
$$
Then, we can introduce the acceleration and the vorticity vector $\omega^\mu$ with
the usual definitions:
\begin{eqnarray*}
A^\mu &= &u \cdot \partial u^\mu  \\
\omega^\mu &=& \frac{1}{2} \epsilon^{\mu\nu\rho\sigma} \partial_\nu u_\rho u_\sigma
\end{eqnarray*}
The antisymmetric part of the tensor $\partial_\mu u_\nu$ can then be expressed as
a function of $A$ and $\omega$:
$$
  \frac{1}{2}\left( \partial_\nu u_\mu - \partial_\mu u_\nu \right) = \frac{1}{2} 
   \left( A_\mu u_\nu - A_\nu u_\mu \right) + \epsilon_{\mu\nu\rho\sigma} \omega^\rho 
  u^\sigma
$$
therafter plugged into the (\ref{basic}) to give:
\begin{eqnarray}\label{spindeco}
S^\mu(x,p)  &=& \frac{1}{8m} (1-n_F)  \epsilon^{\mu\nu\rho\sigma} p_\sigma
   \nabla_\nu (1/T) u_\rho \\
&+&  \frac{1}{8m} (1-n_F) \; 2 \, \frac{\omega^\mu u \cdot p - u^\mu \omega \cdot p}{T} \\
&-& \frac{1}{8m}(1-n_F)\frac{1}{T} \epsilon^{\mu\nu\rho\sigma} p_\sigma  A_\nu u_\rho
\end{eqnarray}
Hence, polarization stems from three contributions: a term proportional to the 
gradient of temperature, a term proportional to the vorticity $\omega$, and a term
proportional to the acceleration. Further insight into the nature of these terms
can be gained by choosing the particle rest frame, where $p=(m,{\bf 0})$ and restoring
the natural units. The eq.~(\ref{spindeco}) then certifies that the spin in the 
rest frame is proportional to the following combination:
\be\label{restframe}
{\bf S}^*(x,p) \propto \frac{\hbar}{KT^2} \gamma {\bf v} \times \nabla T + 
\frac{\hbar}{KT} \gamma (\omegav - (\omegav \cdot {\bf v}) {\bf v}/c^2) 
+ \frac{\hbar}{KT} \gamma {\bf A} \times {\bf v}/c^2
\ee
where $\gamma = 1/\sqrt{1-v^2/c^2}$ and all three-vectors, including vorticity,
acceleration and velocity, are observed in the particle rest frame. 

The three independent contributions are now well discernible in eq.~(\ref{restframe}). 
The second term scales like $\hbar \omega/KT$ and is the one already known from 
non-relativistic physics, proportional to the vorticity vector seen by the particle 
in its motion amid the fluid, with an additional term vanishing in the non-relativistic 
limit. The third term is a purely relativistic one and scales like $\hbar A/KTc^2$;
it is usually overwhelmingly suppressed, except in heavy ion collisions where the 
acceleration of the plasma is huge ($A \sim 10^{30} g$ at the outset of hydrodynamical
stage). The first term, instead, is a new non-relativistic term \cite{becaspin} and 
applies to situations where the velocity field is not parallel to the temperature 
gradient. For ideal uncharged (thus relativistic) fluids, this term is related 
to the acceleration term because the equations of motion reduce to:
$$
 \nabla_\mu T = T A_\mu/c^2
$$
Therefore, being the QGP a quasi-ideal fluid and almost uncharged at very high 
energy, the first and third term are tightly related. It can be shown that they
contribute non-trivially to the final predicted polarization \cite{karpenko}.

%------------------------------------------------------------------------------
\subsection{Open theoretical issues}
%------------------------------------------------------------------------------

As spin physics in relativistic fluids is a completely new subject, much work is still to be 
done and several theoretical questions are still to be answered. For instance, as 
has been mentioned, the formula (\ref{basic}) has been obtained by means of an 
educated {\em ansatz} of the Wigner function of the Dirac field and an exact derivation 
in the framework of quantum field theory is still to be found, even at the global 
thermodynamic equilibium, when the density operator reads:
\be\label{densop2}
  \rho = \frac{1}{Z} \exp \left[ - b_\mu {\wP}^\mu  
  + \frac{1}{2} \varpi_{\mu\nu} \wJ^{\mu\nu} + \zeta \wQ \right]
\ee

A major theoretical progress in this topic is compelling not just because of our
aim at perfection, but in view of the by now demonstrated ability of the experiments
to make differential measurements of polarization. Specifically, new results 
presented in this conference \cite{niida} have shown a disagreement with the 
predictions. Particularly, the azimuthal dependence of the mean spin component 
along the angular momentum $S_J$ is at odds with the prediction of the hydrodynamic 
model \cite{beca2015} (see also \cite{voloshin} and references therein). 
Furthermore, it has been predicted, based again on the hydrodynamic 
model, that the component of the mean spin along the beam axis $S_b$, features 
oscillations in the azimuthal plane similar to the elliptic flow \cite{becakarp}. 
This pattern has been confirmed in a subsequent calculation based on AMPT \cite{xiao}. 
Indeed, these oscillations, with the predicted periodicity, 
have been prelimianarly observed by STAR and reported in this conference \cite{niida}, 
yet with a flipped sign compared to the predictions. Needless to say, was not for 
this mismatch, this finding would be a spectacular confirmation of the model.

Identifying the reason of the aforementioned discrepancies requires a deep theoretical 
investigation.\\ 
The first possibility could be an incorrect choice of the initial hydro
conditions, particularly the initial longitudinal flow profile. This hypothesis
can be tested numerically with a dedicated study with presently available codes.\\
A second possibility is the effect of decays of higher lying states. Indeed, all
theoretical predictions of polarization as a function of momentum only include 
primary particles, that is $\Lambda$ hyperons emitted from the hadronizing hypersurface.
However, most $\Lambda$'s stem from the decays of higher lying states and the 
polarization transfer is a non-trivial function of the momentum of the decay product.
While global polarization transfer in decays producing a $\Lambda$ hyperon has 
been calculated \cite{becalisa} and its impact evaluated numerically \cite{becakarp}, 
nothing is known about it differentially in momentum space. Even if unlikely
at first glance, a sign flip between primary and secondary $\Lambda$'s is 
possible.\\
A third possibility is that local equilibrium of spin degrees of freedom is not
achieved as quickly as momentum and, consequently, the evolution of spin density
is not just driven by thermal vorticity. Thereby, in order to describe a polarized 
relativistic, C-invariant fluid (such as the QGP at very high energy) a spin tensor 
$\spt$ is necessary \cite{becaflork}. 

The latter is an intriguing option as it implies a profound consequence. The need
of a spin tensor entails that different quantum stress-energy tensors are not 
equivalent to describe the fluid \cite{becatinti} and, more importantly, there should be
measurable effects in the predicted polarization pattern. The hydrodynamic
equations of a relativistic polarizeable fluid with particles and antiparticle would
be supplemented by the angular momentum continuity equation:
\be 
 \partial_\lambda \spt^{\lambda,\mu\nu} = T^{\nu\mu} - T^{\mu\nu}
\ee
and that the hydrodynamic variables include also an antisymmetric tensor $\Omega_{\mu\nu}$
\cite{becaflork} playing the role of the thermodynamically conjugate variable of 
the spin tensor, much the same way as the chemical potential is the thermodynamically 
conjugate variable to a conserved charged current. 

It is worth stressing that, with regard to polarization, the case of a non-relativistic 
fluid is in essence different from that of a non-relativistic fluid. In the former,
polarization can be described by magnetization because of the full matter-antimatter
asymmetry. In the latter, this is not enough and if particles and anti-particles
are both polarized in the same direction, magnetization, which is C-odd, cannot 
make the job; a C-even polarization tensor, the spin tensor, is needed.

The transition from the familiar relativistic hydrodynamics to relativistic hydrodynamics 
with spin \cite{florkowski} is a major step forward and would certainly require
a great amount of theoretical and numerical work. QGP offers the unique opportunity 
to test relativistic theories in the laboratory.

%**************************************************************************
\section{Conclusions}
%**************************************************************************

Polarization and Chirality have opened a new window in heavy ion physics. 
From a theory viewpoint, a new outlook, forcing us to rethink the foundations of 
relativistic hydrodynamics and kinetic theory in a fully quantum framework.
The study of the quantum features of Quark Gluon Plasma have exciting 
connections with fundamental physics problems even beyond QCD.

%**************************************************************************
\section*{Acknowledgments}
%**************************************************************************

This work was partially supported by the INFN grant {\em Strongly Interacting Matter}
(SIM). I would like to acknowledge interesting discussions with V. Ambrus, M. Buzzegoli, 
W. Florkowski, B. Friman, X. G. Huang, M. Kaminsky, I. Karpenko, D. Kharzeev, J. F. Liao, M. Lisa, 
E. Speranza, M. Stephanov, L. Tinti, G. Torrieri, I. Upsal, S. Voloshin.
I would like to esepcially thank I. Karpenko for providing me with plots used in 
my presentation.



%% References


%\bibliographystyle{elsarticle-num}



\end{document}